\documentclass[a4paper,ngerman,english,aps,manuscript,aps,prl]{revtex4}
\usepackage[T1]{fontenc}
\usepackage[latin9]{inputenc}
\usepackage{color}
\usepackage{float}
\usepackage{amsmath}
\usepackage{graphicx}



\usepackage{babel}
\begin{document}

\preprint{Translocation of polymer chains through lipid bilayers}

\title{Critical adsorption controls translocation of polymer chains through
lipid bilayers and permeation of solvent}

\author{Jens-Uwe Sommer}

\email{sommer@ipfdd.de}

\affiliation{Leibniz-Institut für Polymerforschung Dresden e.V., Hohe Strasse
6, 01069 Dresden, Germany and Technische Universität Dresden, Institute
of Theoretical Physics, 01069 Dresden, Germany}

\author{Marco Werner}

\email{werner-marco@ipfdd.de}

\affiliation{Leibniz-Institut für Polymerforschung Dresden e.V., Hohe Strasse
6, 01069 Dresden, Germany and Technische Universität Dresden, Institute
of Theoretical Physics, 01069 Dresden, Germany}

\author{Vladimir A. Baulin}

\email{vladimir.baulin@urv.cat}

\affiliation{ICREA, 23 Passeig Lluis Companys, Barcelona 08010, Spain and Departament
d\textquoteright{}Enginyeria Química, Universitat Rovira i Virgili,
26 Av. dels Paisos, Catalans, Tarragona 43007, Spain }
\begin{abstract}
Monte Carlo simulations using an explicit solvent model indicate a
new pathway for translocation of a polymer chain through a lipid bilayer.
We consider a polymer chain composed of repeat units with a given
hydrophobicity and a coarse-grained model of a lipid bilayer in the
self-organized liquid state. By varying the degree of hydrophobicity
the chain undergoes an adsorption transition with respect to the lipid
bilayer. Close to the transition point, at a properly balanced hydrophobicity
of the chain, the membrane becomes transparent with respect to the
chain. At the same time the solvent permeability of the bilayer is
strongly increased in the region close to adsorbed chain. Our results
indicate that the critical point of adsorption of the polymer chain
interacting with the fluctuating lipid bilayer could play a key role
for the translocation of molecules though biological membranes.
\end{abstract}
\maketitle
Lipid bilayers formed by self-organization of amphiphilic molecules
form the natural border of living cells, thus protecting the interior
of the cell from the environment, but also allowing for exchange of
substances and information between the cell and its surroundings.
A challenging problem is to understand the translocation of biopolymers
through lipid bilayers. In particular it would be desirable to know
under which conditions polymers can cross the cell membrane which
otherwise forms a strong barrier against exchange of larger molecules.
Some anti-microbial peptides~\cite{Brogden_NatRevMicrobiology05},
cell-penetrating peptides~\cite{Nekhotiaeva_FASEB04} and drug-delivery
polymers~\cite{Lynch_Biomat10} have been shown to penetrate through
phospholipid bilayers. Despite considerable effort devoted to the
problem, physical principles underlying their membrane activity and
translocation behavior are not well understood. The suggested mechanisms
of interaction of biopolymers with bilayers~\cite{Mathot_JCR07}
imply formation of static structures in the membranes such as stable
pores. However, the formation of static well-ordered structures by
polymers without a secondary structure is questionable, and their
translocation through membranes~\cite{Lynch_Biomat10,Goda_Biomaterials10}
 may indicate the existence of other translocation mechanisms. Recent
studies~\cite{Lynch_Biomat10} have shown that at certain conditions
amphiphilic copolymers can translocate through membranes and simultaneously
enhance permeability of membranes for low molecular components without
evidence of formation of static pores or other structures. Such ability
to facilitate the permeation of membranes has important applications
in red blood cells preservation~\cite{Lynch_Biomat10}, trans-membrane
drug delivery~\cite{Mathot_JCR07}, enhancement of intestinal wall
for drugs~\cite{Muranashi_CR90} and gene delivery~\cite{Pack_NRDD05}.

Understanding these mechanisms of interaction of polymers with bilayers
is of particular importance for development of efficient delivery
vectors which are able to transport drugs or foreign DNA into the
cell~\cite{Prochiantz_ADDR08,Prochiantz_NATMETH07,Zorko_ADDR05}.
On the other hand, there are indications that biopolymers interacting
with cell membranes might significantly increase permeability with
respect to small molecules such as trehalose depending on the amount
of charged side groups of the polymer \cite{Lynch_Biomat10}.

In this work we explore a simple model-system where the lipid bilayer
is formed by coarse grained and flexible amphiphilic molecules using
an explicit solvent model to prevent artifacts due to strong hydrophobic
interactions, but also to be able to consider the permeability of
the self-organized membrane with respect to the solvent. Polymer chains
interact with the membrane via an effective hydrophobicity of the
monomers which should mimic the effect of a hetero-polymer having
both hydrophilic and hydrophobic components. Charge effects will be
disregarded in our work focusing on short-range interactions between
the monomers of the chains and the monomers of the lipid-molecules.
These simplifications allow for the understanding of a fundamental
relation between the hydrophobic interactions between polymers and
lipids and the translocation and permeability barrier of the lipid-bilayer.
In particular we will show that lipid-bilayers become transparent
for the polymer chains if the effective hydrophobic interactions drives
the chains near the adsorption threshold with respect to the membrane.
Deviating from this adsorption threshold only slightly the chain is
either rejected from the membrane, or trapped inside the hydrophobic
core. We also demonstrate that close to the adsorption threshold the
permeability of the membrane with respect to the solvent molecules
in the environment of the chain is strongly increased. After presenting
briefly some details of the simulation method we will report our results.

\begin{figure}
\includegraphics{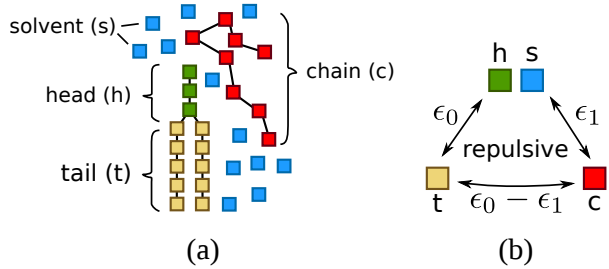}

\caption{Implementation of the bond fluctuation model. (a) coarse grained representation
of lipids, explicit solvent and polymer chains as used in our BFM
simulations. (b) repulsive interactions between the four components
as defined in (a) and their associated energy constants $\epsilon_{0}$
and $\epsilon_{1}$. \label{fig:ImplementBFM}}
\end{figure}
In order to simulate lipid bilayers and their interaction with linear
chains we use the bond fluctuation method (BFM)~\cite{bfm,deutsch_JCP90},
using a short-range interaction model between the different species.
The model is well established to simulate various polymer systems,
details of our implementation of the interaction model can be found
in Ref.~\cite{werner_EPJE10}. In this work we consider an explicit
solvent model which avoids freezing effects and related artifacts
due to strong attractive interactions imposed by solvent-free models~\foreignlanguage{ngerman}{\cite{ReddyYethiraj_MM06,luettmerBinder_JCP08}}.
The essential details of the model are illustrated in Fig.\ref{fig:ImplementBFM}.
We simulate two-tailed lipids with five monomers per tail (t) and
three monomers per head (h) as well as single polymer chains (c) of
various chain lengths, $L_{c}$, see Fig.~\ref{fig:ImplementBFM}(a).
Furthermore, an explicit solvent (s) represented by unconnected BFM-units
is taken into account. The simulation box is filled up by solvent
reaching a volume occupation of $0.5$, which corresponds to a dense
system in the framework of the BFM. Within the framework
of the simulation model packing effects of the lipids are taken into
account. Since we use local monomer moves only the algorithm mimics
a proper diffusional dynamics of polymers, lipids and solvent.

\begin{figure}
\includegraphics[width=8cm]{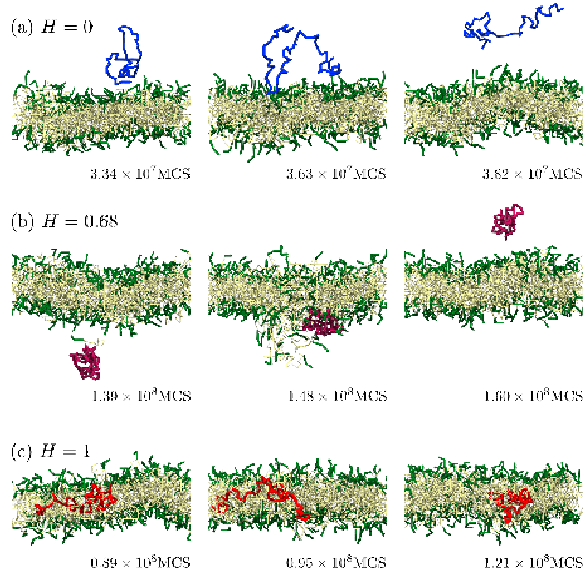}\caption{Snapshots of BFM-simulations (see also Fig.~\ref{fig:ImplementBFM}
and Eq.(\ref{eq:Def_epsilon})) of lipid bilayers and linear polymer
chains ($L_{C}$ = 64) with various relative hydrophobicity, $H$,
taken at given simulation times. \label{fig:Snapshots-of-BFM-simulations}}
\end{figure}

The hydrophobic effect is modeled as a short-range repulsive interaction
(energy penalty for pair contacts) between monomers of different hydrophobicity.
Each of the $24$ next nearest lattice sites of a tagged BFM-monomer
of type $A$ is associated with an energy penalty, $\epsilon_{AB}$,
if it is occupied by an interacting monomer of type $B$ \cite{hoffmann_JCP97}.
The resulting internal energy of the tagged monomer rules the elementary
Metropolis step. In a four component system (t,h,c,s) in the most
general case six interaction constants can be fixed. We used a simplified
interaction model according to Fig.~\ref{fig:ImplementBFM}, where
only two independent interaction energies, $\epsilon_{0}$ and $\epsilon_{1}$
are considered according to
\begin{align}
\epsilon_{0}= & \,\epsilon_{t,h}=\epsilon_{t,s}\nonumber \\
\epsilon_{1}= & \,\epsilon_{h,c}=\epsilon_{s,c}=\epsilon_{0}-\epsilon_{t,c}\label{eq:Def_epsilon}\\
\epsilon_{h,s}= & \,0\,\,.\nonumber
\end{align}
Hence, we assume that solvent (s) and heads (h) are indistinguishable
with respect to their interactions with other (hydrophobic) species.
Applying the above model to random starting configurations of lipids
in solvent (without chains), self-assembled lipid bilayers in a planar-
or vesicular shape were obtained depending on boundary conditions
and lipid volume fraction. In the following we consider planar bilayers
consisting of $300$ lipids in a cubic box of size $64$ with periodic
boundary conditions in all directions. The starting configuration
of the bilayer has been chosen such that the self-organized membrane
spreads over the periodic boundaries in $x$- and $y$-direction and
shows a stable orientation perpendicular to the $z$-axis. Therefore,
the $(x,y)-$ plane of the lattice is subsequently used as a fixed
reference-plane for the bilayer. We have chosen $\epsilon_{0}=0.8$
(units of $k_{B}T$) which leads to the formation of stable bilayers.
To characterize the extension of the bilayer, we calculated the center
of mass $\bar{z}=\langle z\rangle$ and the variance $\sigma_{z}{}^{2}=\langle z^{2}\rangle-\bar{z}^{2}$
of the $z$-positions of all tail monomers during our simulations.
We define the boundaries of the bilayer by $z_{\pm}=\bar{z}\pm2\sigma_{z}$.
We added single chains of various chain length, $L_{c}$, to the simulation
box with an effective hydrophobicity, $\epsilon_{1}$. Typical snapshots
for $L_{c}=64$ are shown in Fig.~\ref{fig:Snapshots-of-BFM-simulations}.

In the following we show simulation results under variation of $\epsilon_{1}$
at constant $\epsilon_{0}=0.8$. We define the \emph{relative hydrophobicity}
of the chain according to
\begin{equation}
H=\epsilon_{1}/\epsilon_{0}\,\,.\label{eq:Def_h}
\end{equation}

As can be seen in Fig.~\ref{fig:Snapshots-of-BFM-simulations}(a),
for $H=0$ the (hydrophilic) chain is rejected by the lipid bilayer
and its conformations correspond to a self-avoiding walk in good solvent.
For $H=1$ (hydrophobic) the chain is collapsed in the solvent but
is strongly attracted and eventually absorbed by the bilayer's core.
At intermediate values close to $H=0.68$ (partial hydrophobic) the
chain can be found in the solvent phase as well as penetrating the
bilayer . Note that the chain is collapsed at this value of $H$
and explicit solvent simulation enables a non-frozen liquid-like dynamics
of the chain.

\begin{figure}
\includegraphics[width=6cm]{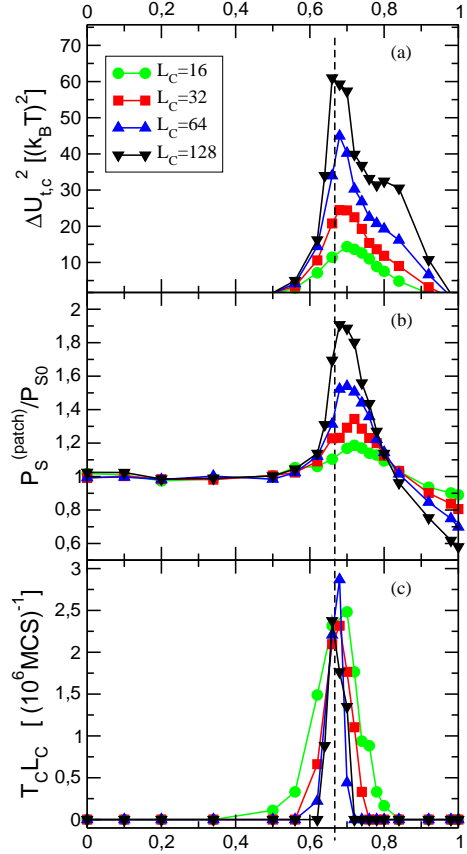}

\caption{$\mathbf{(a)}$ Fluctuation of the internal energy, $\Delta U_{\mathrm{t,c}}{}^{2}=\langle U_{\mathrm{t,c}}^{2}\rangle-\langle U_{\mathrm{t,c}}\rangle^{2}$
, corresponding to nearest neighbor-contacts between tail-monomers
in a lipid bilayer and monomers of a linear chain of length $L_{c}$
depending on relative hydrophobicity, $H$, see Eq. (\ref{eq:Def_epsilon}).
The extrapolated adsorption transition, $H_{A}=0.667\pm0.005$, is
indicated by the vertical dashed line.\label{fig:Rescaled-fluctuations-U}$\mathbf{(b)}$
Simulation results for the solvent-permeability, $P_{S}^{(patch)}$
, of a circular membrane patch (projected radius $\approx15$, see
text) around the projected center of mass of a polymer chain depending
on the relative hydrophobicity $H$. The results are normalized to
the permeability for an membrane not interacting with polymer chains
(corresponds practically to $H=0$), $P_{S,0}$. $\mathbf{(c)}$ Simulation
result for the rescaled frequency of translocations, $T_{C}$, of
a polymer chain of length $L_{c}$ through a lipid bilayer depending
on relative hydrophobicity of the chain.\label{fig:U_tc_fluct_and_Ps_and_Tc}}
\end{figure}

By increasing the relative hydrophobicity the bilayer changes its
energetic characteristics from a potential barrier for $H=0$ into
a potential trap for $H=1$ with respect to the chain. This resembles
an adsorption scenario of a chain at a penetrable potential well.

In general, one expects a characteristic value of $H_{A}$ where the
chain is just at the transition to the adsorbed state. In order to
characterize the transition point in our model we have calculated
the fluctuation of the contact energy $U_{\mathrm{t,c}}$, (heat capacity)
between tail monomers and chain monomers as shown in Fig.~\ref{fig:U_tc_fluct_and_Ps_and_Tc}(a).
At the transition point the fluctuations display a peak which should
approach the critical point of adsorption for $L_{c}\to\infty$~\cite{ekb_82}.
For finite chains adsorption corresponds to a crossover scenario where
the effective adsorption threshold is increased for decreasing chain
length according to $(H-H_{A})/H_{A}\sim N^{-\phi}$, where $\phi$
denotes the crossover-exponent for polymer adsorption, whose values
depend on the problem at hand~\cite{ekb_82,bd}. Our results suggest
that the adsorption transition is close to $H_{A}\simeq0.667\pm0.005$.
We note that a value of $H_{A}>0.5$ indicates a higher chemical potential
of the bilayer's core with respect to the solvent (higher density,
perturbation of the lipid conformation and local fluctuations of the
bilayer due to embedding of the chain or chain parts).

Let us consider the frequency of permeations (translocations) of molecules
of species $\alpha$ through the model bilayer,
\begin{equation}
T_{\alpha}=\frac{n_{\alpha}}{\Delta t}\quad\text{and}\,\,\, P_{\alpha}=\frac{T_{\alpha}}{\mathcal{A}}\,\,,\label{eq:Def_T}
\end{equation}
where $n_{\alpha}$ is the number of permeation events and $\Delta t$
is the simulation time in units $10^{7}$Monte Carlo steps (MCS).
The permeability, $P_{\alpha}$, of the membrane with respect to species
$\alpha$ is proportional to the translocation frequency $\text{T}_{\alpha}$
in the case of a given concentration difference on both sides of the
membrane. For simplicity, let us here define the permeability as a
translocation frequency per surface area where the surface area, $\mathcal{A}$,
is given by the projection of the membrane to the $(x,y)$-plane of
the lattice.

We detected permeation events both of solvent molecules as well as
of the linear chain using the trajectories of their centers of mass
with a time resolution of $100$~MCS. The $z$-components of the
trajectories have been analyzed with respect to their crossing points
with two thresholds, $l_{\pm}$, on both sides of the bilayer center,
$\bar{z}$. If a molecule passed successively through both planes
defined by $l_{+}$ and $l_{-}$ we counted one permeation event for
the respective molecule species. For the solvent molecules we define
the two thresholds, $l_{\pm}$, to be identical to the boundaries
of the bilayer, $l_{\pm}^{(S)}=z_{\pm}$. For the polymer chain we
fixed the thresholds to $l_{\pm}^{(C)}=\bar{z}\pm23$, where we concluded
from the center of mass distribution of the longest chains, $L_{C}=128$,
that they are well separated from the bilayer.

The results for the frequency of translocations of the chain, $T_{C}(H)$,
are displayed in Fig.~\ref{fig:U_tc_fluct_and_Ps_and_Tc}(c). Here,
we have rescaled the frequency of translocations with the center of
mass diffusivity of the free chain which is $\sim1/L_{C}$. Translocation
of the chain is clearly shown close to the adsorption transition.
We note that at $H_{0}=1/2$ the lipid tails are equally repulsive
as the solvent with respect to the chains. However, in the self-assembled
state the tails are stretched and partially ordered which imposes
a Free energy effort to insert the chain into the bilayer at $H_{0}$
similar to the case of a polymer brush~\cite{SolisTang_MM10,Halperin_LAN07}.
Let us denote this Free Energy barrier by $\Delta F_{ins}=N\Delta\mu$.
In order to compensate this barrier an effective attraction strength
of $\epsilon=\chi_{0}\epsilon_{0}\Delta$ with $\Delta=(H-H_{0})/H_{0}$
is necessary. Here, $\chi_{0}$ accounts for the average number of
monomers interacting with a given chain monomer. For $\mu=\epsilon$
the free energy barrier vanishes. On the other hand, surface adsorption
of the chain can occur, because for monomers directly in contact with
the bilayer's core an effective attraction potential with a depth
of $-\epsilon/\chi_{0}$ is forming on both sides of the core. The
chain can be localized at this potential trap with a limited penetration
into the core. A closer analysis for the case of an ideal chain reveals
an adsorption threshold~\cite{sbs} which very close to $\mu=\epsilon$.
This can explain the coincidence between the critical point of adsorption
and the translocation of the chain. However, we note that the ideal
chain model might not be fully appropriate here since the chain is
in a collapsed state at the transition point.

An interesting measure for perturbations of the bilayer due to an
adsorbing polymer chain is the permeability, $P_{S}$, of the bilayer
with respect to the solvent molecules. For each solvent permeation
event we have recorded the distance between the position of the solvent
unit and the center of mass of the chain projected into the reference
plane of the bilayer, $\vec{d}$, taken in the moment of the beginning
of the translocation event. In Fig.~\ref{fig:U_tc_fluct_and_Ps_and_Tc}(b)
we show simulation results for the local permeability $P_{S}^{(patch)}$
through a circular membrane patch of (projected) radius $|\vec{d}|<15$.
This defines a patch on the membrane that follows the center of mass
of the chain as a ``shadow''. For chains length $L_{C}=128$ the
local permeability $P_{S}^{(patch)}$ increases up to a factor of
almost two with respect to the unperturbed bilayer if the chain is
close the adsorption transition. It is notable that for strong hydrophobicity
of the chain, $H\to1$, the permeability decreases up to a factor
of almost two, indicating a stabilizing effect of the embedded chain.
The peak of the solvent permeability correlates with the effective
adsorption transition of the chain for different chain lengths.

\begin{figure}[H]
\includegraphics[width=8cm]{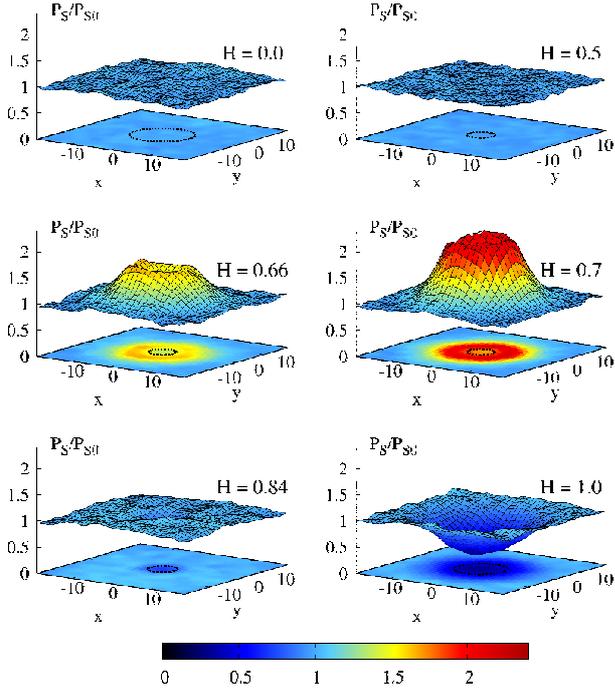}

\caption{Local solvent-permeability $P_{S}$ of a lipid bilayer as a function
of the Cartesian distance vector $(x,y)$ in the membrane's reference
plane from the projected center of mass position of a single polymer
chain. We show results for various degrees of the relative hydrophobicity,
$H,$ normalized to the average permeability $P_{S0}$ in the case
of $H=0$.\label{fig:Solvent-permeability}}
\end{figure}

In Fig.~\ref{fig:Solvent-permeability} we show the local permeability
$P_{S}(\vec{d)}$ for various hydrophobicities $H$ as a function
of the projected distance vector between entering points and chain,
$\vec{d}.$ The permeability of the solvent is only increased in the
immediate environment of the adsorbing chain near $H_{A}.$ Note that
close to the center of the chain solvent permeability is partially
screened by the polymer globule.

To conclude our results we have shown that for a simplified model
of the lipid bilayer the point of adsorption of the chain controls
both the solvent permeability and the chain translocation through
the membrane. This can be explained by assuming the membrane core
as a potential well for the polymer chain which becomes transparent
if the relative hydrophobicity of the monomers is balanced in such
a way that both the solvent and the hydrophobic core of the membrane
are equally poor environments for the chain. The translocation maximum
and the point of adsorption indicate a threshold for insertion of
chain monomers into the bilayer core. Perturbation of the membrane
properties due to chain adsorption correlates with enhanced solvent-permeability
of the membrane close to the adsorption point. In biological systems
the polymers are usually represented by amphiphilic
copolymers, such as polypeptides, with a possibly disordered arrangement
of both hydrophilic and hydrophobic monomers. Hydrophobic compensation
and adsorption discussed in our work then depends on the ratio of
hydrophilic and hydrophobic monomers. Thus, a common feature of membrane-crossing
polymers can be the proper balance of the effective hydrophobicity.

\medskip{}



\end{document}